\documentclass[11pt,a4paper]{article}

\pdfoutput=1
\usepackage{jheppub}

\usepackage{amssymb}

\usepackage{braket}
\usepackage{multirow}
\usepackage{graphicx}
\usepackage{wrapfig,enumerate,slashed}

\usepackage{footmisc}
\usepackage{amsmath}
\usepackage{wasysym} 
\usepackage{color}
\usepackage{xcolor}
\usepackage{comment}
\usepackage{siunitx}
\usepackage{booktabs}

\DeclareMathAlphabet{\mathbold}{U}{zeur}{b}{n}

\usepackage[labelfont=bf,font={sl,small}]{caption}

\renewcommand\({\left(}
\renewcommand\){\right)}
\renewcommand\[{\left[}
\renewcommand\]{\right]}

\def\beq{\begin{equation}}
\def\eeq{\end{equation}}
\def\[{\begin{equation}}
\def\]{\end{equation}}

\newcommand\tr{\operatorname{tr}}
\newcommand\OSk{\mathcal{O}_{\text{Sk}}}

\newcommand\im{\operatorname{Im}}
\newcommand\hc{\text{h.c.}}

\begin{document}
\numberwithin{equation}{section}

\title{The Emergence of Electroweak Skyrmions through Higgs Bosons}

\author{Juan Carlos Criado, Valentin V. Khoze and Michael Spannowsky}

\affiliation{Institute for Particle Physics Phenomenology, Department of Physics, Durham University, Durham, DH1 3LE, UK}

\emailAdd{juan.c.criado@durham.ac.uk}
\emailAdd{valya.khoze@durham.ac.uk}
\emailAdd{michael.spannowsky@durham.ac.uk}

\abstract{
Skyrmions are extended field configurations, initially proposed to describe baryons as topological solitons in an effective field theory of mesons. We investigate and confirm the existence of skyrmions within the electroweak sector of the Standard Model and study their properties. We find that the interplay of the electroweak sector with a dynamical Higgs field and the Skyrme term leads to a non-trivial vacuum structure with the skyrmion and perturbative vacuum sectors separated by a finite energy barrier. We identify dimension-8 operators that stabilise the electroweak skyrmion as a spatially localised soliton field configuration with finite size. Such operators are induced generically by a wide class of UV models. To calculate the skyrmion energy and radius we use a neural network method. Electroweak skyrmions are non-topological solitons but are exponentially long lived, and we find that the electroweak skyrmion is a viable dark matter candidate. While the skyrmion production cross section at collider experiments is suppressed, measuring the size of the Skyrme term in multi-Higgs-production processes at high-energy colliders is a promising avenue to probe the existence of electroweak skyrmions.}


\preprint{IPPP/20/64}

\maketitle


\section{Introduction}

The discovery of the Higgs boson \cite{Aad:2012tfa,Chatrchyan:2012ufa}, a remnant of the mechanism that spontaneously breaks the electroweak symmetry of the Standard Model, firmly established the existence of a scalar sector at the electroweak scale. Higgs interactions with itself and other particles has profound consequences in particle theory for the structure of the scalar potential and of the electroweak vacuum. It is also well-known that the concept of spontaneous symmetry breaking of gauge theories was first observed and described in superconductive condensed matter systems \cite{Anderson:1963pc}, before being applied to quantum field theories relevant to high-energy physics \cite{Higgs:1964pj, Englert:1964et, Guralnik:1964eu}. 

Skyrmions are energetically stable static field configurations that can describe new particle degrees of freedom in QFT and have at least as long history as the Higgs boson itself. Skyrmions were first introduced in \cite{Skyrme:1961vq} to represent baryons as topological solitons emerging in an effective field theory of mesons~\cite{Skyrme:1961vq,Witten:1979kh}. To ensure that the soliton is energetically stable, a higher-derivative operator, the so-called Skyrme term, was introduced and added to the non-linear sigma model of mesons. Inclusion of such a term in the effective Lagrangian effectively bypasses Derrick's theorem \cite{Derrick:1964ww} and allows the skyrmion to be realised as a spatially localised soliton field configuration with finite size. In this context, as the low-energy description of nucleons in strong interactions, the Skyrme model predictions are within $30\%$ of experimental values~\cite{Adkins:1983ya,Zahed:1986qz}, but there have been no direct experimental evidence for skyrmions in particle physics.\footnote{Of course, this does not contradict the original statement that in strong interactions skyrmions  provide a correct description of qualitative features of baryons in the effective meson theory below the confinement scale of QCD.}

This is to be contrasted with condensed matter systems where skyrmion-like field configurations have been observed.
As vortex-like textures of magnetic moments, skyrmions were experimentally observed and studied in magnetically ordered materials \cite{bog89,bog94,roes06,leon16}, for a review see \cite{Lanc19}. Due to their small size and long lifetime these magnetic skyrmions have received much attention as a promising avenue to facilitate low-energy magnetic data storage devices. Thus, skyrmions are realised in nature and can be studied in a range of condensed matter systems {\it in vivo}.  

Motivated by the striking parallels between the Higgs boson and skyrmion histories in particle physics and in condensed matter, we would like to reexamine skyrmions in particle physics, and in particular, skyrmions in the electroweak theory of the Standard Model.

Electroweak skrymions have been studied previously assuming an infinitely heavy non-dynamical Higgs field \cite{Ambjorn:1984bb,Eilam:1985tg,Brihaye:1989ej,Carson:1990yk,Farhi:1995aq,Ellis:2012bz,Ellis:2012cs}, but, interestingly, despite the discovery of the Higgs boson, to our knowledge, they have never been studied in light of the electroweak vacuum and scalar potential as realised in the Standard Model and its extensions. Thus, the profound consequences a dynamical Higgs field has on the properties of electroweak skyrmions have not been investigated so far. This is important for two reasons. First, in the presence of a dynamical Higgs field the candidate skyrmion field configuration could rapidly unwind and decay into elementary Higgs bosons. 
We find that electroweak skyrmions with masses $M_{\text{Sk}} \lesssim  \SI{10}{TeV}$
 readily co-exist with dynamical Higgs and gauge fields as non-topological solitions.    
Secondly, if one takes the lowest-order 4-derivative operator needed to stabilise the skyrmion in the non-dynamical-Higgs case, and naively introduces a dynamical Higgs, one gets a dimension-12 operator. However, we find that skyrmion-stabilising operators appear already at dimension 8 of the perturbative series of the effective theory, e.g. for Standard Model particle content and symmetries the so-called Standard Model Effective Field Theory framework (SMEFT) \cite{Grzadkowski:2010es}. In turn, the possible existence of an electroweak skyrmion and its phenomenological implications to collider phenomenology, dark matter and early Universe physics underline the importance of including dimension-8 operators in global EFT analyses.

In Sec.~\ref{sec:theory} we outline how electroweak skyrmions arise and can be energetically stabilised in the presence of a dynamical Higgs field. This includes a discussion of stabilising dimension-8 operators and how skyrmion production and decay relate to B+L violating processes in the Standard Model. In Sec.~\ref{sec:sum} we calculate the skyrmion energy using novel machine-learning techniques. That the skyrmion-stabilising dimension-8 operators are induced rather generically in a wide range of Standard Model extensions we show in Sec.~\ref{sec:ext}. In Sec.~\ref{sec:pheno} we give a brief overview on the possibility to study skyrmions or the operators required to stabilise them at current and future colliders. We note that electroweak skyrmions can be viable dark matter candidates. Finally in Sec.~\ref{sec:conclusions} we offer a summary and conclusions.

\section{Electroweak Skyrmion theory}
\label{sec:theory}

\subsection{Preliminaries}
\label{sec:preliminaries}

We start with the Lagrangian for the SM Higgs scalar coupled to the $SU(2)_L$ gauge fields,
\[
{\cal L}\,=\, -\frac{1}{2g^2}\,{\rm tr} \, W^{\mu \nu}W_{\mu\nu} \,+\,
\frac{1}{2} {\rm tr} \left(\left(D^\mu  \Phi\right)^\dagger D_\mu \Phi\right)\,-\,
\frac{\lambda}{4} \left({\rm tr} \, \left(\Phi^\dagger \Phi\right)-v^2\right)^2\,.
\label{eq:Lsm}
\]
The gauge fields are written in the usual matrix notation $W_\mu=g W_\mu^a \, \tau^a/2$ where $\tau^a$ are the Pauli matrices, and our normalisation for the gauge fields includes the coupling constant $g$. Hence the canonically normalised kinetic term for the gauge fields is $\propto 1/g^2$, with the 
field strength and the covariant derivatives given by,
\[
W_{\mu \nu} \,=\, \partial_\mu W_\nu -\partial_\nu W_\mu-i[W_\mu, W_\nu]\,, \quad
D_\mu \Phi \,=\, (\partial_\mu -iW_\mu) \Phi\,.
\label{eq:FD}
\]
The SM Higgs doublet we choose to write in an equivalent form, as a unitary two by two matrix of complex scalar fields $\phi_0(x)$ and $\phi_1(x)$,
\[
\Phi(x)\,=\, \begin{pmatrix} \phi_0^* & \phi_1 \cr -\phi_1^* & \phi_0 \end{pmatrix},
\label{eq:Phidef}
\]
which can also be readily decomposed into the real scalar field $s(x)$ times the $SU(2)$ matrix field $U(x)$,
\[
\Phi (x) \,=\, s(x) \, U(x) \,, \quad {\rm where} \quad U(x) \in SU(2)\,, \quad s(x) \in R\,.
\label{eq:Udef}
\]
The neutral Higgs field of the SM is obtained by shifting $s(x)$ by its vacuum expectation value,
\[
s(x) \,=\, \frac{1}{\sqrt{2}}(v\,+\, h(x))\,, \quad {\rm with}  \quad m_h\,=\, 125\,{\rm GeV}\,.
\label{eq:h0}
\]
Our choice of the slightly unusual $2\times 2$ matrix conventions for the SM Higgs in \eqref{eq:Phidef}-\eqref{eq:Udef} is dictated, following  \cite{Farhi:1995aq}, by the simplicity of its connection  to  electroweak skyrmions. As we will explain below, the skyrmion solution of the SM Lagrangian with an additional EFT 4-derivative operator, is naturally described in terms of the $SU(2)$ matrix field $U(x)$ in \eqref{eq:Udef}. 
It is straightforward to switch from the notation \eqref{eq:Phidef} to the conventional SM and EFT notation in terms of the complex Higgs doublet
\[
\phi(x) \,=\, \begin{pmatrix} \phi_1 \cr  \phi_0 \end{pmatrix},
\label{eq:phi_norm}
\]
using the representation $\Phi(x)\,=\, \left(\tilde{\phi}, \phi\right)$, where we have defined 
$\tilde{\phi} = \epsilon \cdot \phi^*(x)$. For more detail, we refer to sections~\ref{sec:sum}.  

\medskip

The Lagrangian \eqref{eq:Lsm} describes the bosonic sector of weak interactions of the SM, it is invariant under $SU(2)_L $ 
transformations,\footnote{The {\it global} symmetry is the $SU(2)_L \times SU(2)_R$ transformations of the scalar field $\Phi \to {\cal U}_L\, \Phi \, {\cal U}_R$.}
\[
\Phi(x)  \to {\cal U}_L(x)\,  \Phi(x)\,, \qquad
W_\mu(x)\, \to \, {\cal U}_L(x)\left(W_\mu +i\partial_\mu\right) {\cal U}_L^\dagger(x)\,.
\label{eq:gaugedef}
\]

In order to look for topological and non-topological solitons and other vacuum configurations
of the theory we impose the requirement of finiteness of the energy,
\[
E_{H,W} \,=\, 
\int d^3 x \, \left\{-\frac{1}{2g^2}\,{\rm tr} \, W_{i j} W_{ij} \,+\,
\frac{1}{2} {\rm tr} \left(\left(D_i  \Phi\right)^\dagger D_i \Phi\right)\,+\,
\frac{\lambda}{4} \left({\rm tr} \, \left(\Phi^\dagger \Phi\right)-v^2\right)^2\right\}\, <\, \infty\,,
\label{eq:EHW}
\]
computed in the $W_0=0$ gauge on the static field configurations $W_i({\bf x})$ and $\Phi({\bf x})$ i.e. the fields at fixed $t$ which can be taken $t=0$.
The requirement of $E_{H,W} <\infty$ implies that, as $|\mathbf{x}| \to \infty$ the gauge fields should approach a pure gauge configuration, while $s(x)$ and $U(x)$ should go respectively to its vev and a constant $SU(2)$ matrix $U_\infty$. One can then always choose a gauge in which the following boundary conditions are satisfied~\cite{Spannowsky:2016ile}:
\begin{equation}
  \lim_{|\mathbf{x}| \to \infty} W_i(x) = 0, \qquad
  \lim_{|\mathbf{x}| \to \infty} s(x) = v / \sqrt{2}, \qquad
  \lim_{|\mathbf{x}| \to \infty} U(x) = 1_{2 \times 2}.
  \label{eq:bc-topological}
\end{equation}
Since all the fields are single-valued at spatial infinity, $R^3$ can be compactified in this setting to $S^3 \cong R^3 \cup \{\infty\}$.

The topology of a field configuration can be partly characterized using the Higgs winding number $n_H$ and the Chern-Simons number $n_{\text{CS}}$, given by
\begin{align}
  \label{eq:nHdef}
  n_H
  &= \frac{1}{24 \pi^2} \epsilon_{ijk} \int d^3x
    \tr{\left[
    (U^\dagger \partial_i U) (U^\dagger \partial_j U) (U^\dagger \partial_k U)
    \right]},
  \\
    \label{eq:nCSdef}
  n_{\text{CS}}
  &= \frac{1}{16 \pi^2} \epsilon_{ijk} \int d^3x
    \tr{\left[W_i W_{jk} + \frac{2i}{3} W_i W_j W_k\right]}.
\end{align}
The finiteness of the energy integral in \eqref{eq:EHW} requires that
the $\Phi$ is continuous,  otherwise the derivative term on the r.h.s. of \eqref{eq:EHW} would result in delta functions giving an infinite contribution to the integral.
Then, if $s(\mathbf{x})$ does not vanish for any $\mathbf{x}$, $U$ must be continuous, implying that $n_H$ is a finite integer number, which characterizes the homotopy class of $U$:
\begin{equation}
  U : S^3 \to SU(2) \cong S^3, \qquad \pi_3(S^3) = \{n_H\} = \mathbb{Z}.  
\end{equation}
On the other hand, $n_{\text{CS}}$ does not need to be an integer for a general field $W$.
However, when $W$ is a pure gauge $W_i(x) =\,i \,\mathcal{U}(x)^{-1} \partial_i \,\mathcal{U}(x)$ for some $SU(2)$-valued function $\mathcal{U}$,
\begin{equation}
  n_{\text{CS}} = \frac{1}{24 \pi^2} \epsilon_{ijk} \int d^3x
  \tr{\left[(\mathcal{U}^{-1} \partial_i \,\mathcal{U}) (\mathcal{U}^{-1} \partial_j\, \mathcal{U}) (\mathcal{U}^{-1} \partial_k\, \mathcal{U})\right]}
  \label{eq:nCS-pure-gauge}
\end{equation}
is an integer characterising the homotopy class of $\mathcal{U}: S^3 \to S^3$.

The fact that $n_H$ is a homotopy invariant for continuous $U$ does not
guarantee or even imply the existence of solitons in our model.
In fact, there are three independent reasons for why topological solitons%
\footnote{%
  Topological solitons here refer to extended particles with non-zero finite
  mass (energy) that are protected by the topological conservation law -- their
  charge given by the winding number $n$ is strictly conserved. As a result
  there is an infinite energy barrier separating topological solitons of
  different charges from each other and from the perturbative vacuum.%
} %
do not exist in the weak sector of the SM described by the
Lagrangian~\eqref{eq:Lsm}.  We will now list and then address these reasons in
turn:

\begin{enumerate}
\item Fluctuations of dynamical $s(x)$ field (i.e. interactions with the SM Higgs);
\item Derrick's theorem;
\item Presence of the $SU(2)_L$ gauge fields $W_\mu(x)$.
\end{enumerate}

\noindent The existence of a dynamical degree of freedom $s$ invalidates the assumption of continuity of the $SU(2)$ field $U$ 
on which the topological conservation of the winding number $n_H$ is based.
In other words, the $SU(2)$-valued $U$ field cannot pass through zero, but the $s({\bf x})$ singlet can. When this happens,
$\Phi({\bf x}) = 0 $ for some $\mathbf{x}$ and $n_H$ can safely unwind and change its value without resulting in infinite energy.
Only if $s(\textbf{x})$ was frozen at its expectation value $v$, the zeros of $s$ would be impossible. But for the neutral Higgs field with a finite mass,
in \eqref{eq:h0}, the zeros of $s$ are possible and the winding number is no longer a conserved quantity. At best there are only finite energy barriers 
separating $U$-fields with different values of $n_H$.

\medskip
\subsection{Skyrme term as a dimension-8 SMEFT operator}
\label{sec:2.2}
\medskip

The consequence of the Derrick's theorem is that even a non-topological soliton is impossible in the model~\eqref{eq:Lsm},
unless we add an appropriate higher-dimensional term to stabilise the soliton size~\cite{Skyrme:1961vq}. The leading-order term capable of stabilising the skyrmion solution is,
the dimension-8 operator,
\[
{\cal L}_{\rm \, S}\,=\, 
\frac{1}{8\Lambda^4}\,
 {\rm tr} \left((D_\mu\Phi)^\dagger D_\nu \Phi\,-\, (D_\nu\Phi)^\dagger D_\mu \Phi\right)^2.
\label{eq:stab8}
\]

In the idealised regime of an infinitely heavy Higgs field, the $s(x)$ field is frozen at $s \to v/\sqrt{2}$ and in this limit the 
effective operator above reduces to the corresponding four-derivative term in the gauged non-linear sigma model. If we now also decouple the gauge fields in the covariant derivatives in \eqref{eq:stab8}, we find,
\[
{\cal L}_{\rm \, S} \,\to\, 
\frac{v^4}{32\Lambda^4}\,
 {\rm tr} \left((\partial_\mu U)^\dagger \partial_\nu U\,-\, (\partial_\nu U)^\dagger \partial_\mu U\right)^2.
\label{eq:stab8vA}
\]
Since  $U^\dagger U=1$, we have $U^\dagger \partial_\mu U= - (\partial_\mu U^\dagger) U$ and use this to represent the
expression in brackets in \eqref{eq:stab8vA} in the form of a commutator, as follows,
\[
(\partial_\mu U^\dagger)(\partial_\nu U) -(\partial_\nu U^\dagger)(\partial_\mu U)
\,=\,-\left[U^\dagger D_\mu U\,,\,  U^\dagger D_\nu U \right]
\,.
\label{eq:Skyrmecom}
\]
This expression on the right hand side of \eqref{eq:stab8vA} then takes the form
\[
\frac{1}{32\, e^2}\, {\rm tr} \left[U^\dagger \partial_\mu U\,,\,  U^\dagger \partial_\nu U \right]^2,
\label{eq:SkyrmeT}
\]
which is recognised as the famous 4-derivative commutator squared Skyrme term~\cite{Skyrme:1961vq},
that allows one to bypass the Derrick's theorem and to stabilise the skyrmion solution.
We used here the standard in the skyrmion literature convention~\cite{Adkins:1983ya} for the normalisation factor $\frac{1}{32\, e^2}$ in front of the Skyrme term.
In our setting, it is related to the Wilson coefficient in front of the EFT operator in \eqref{eq:stab8} via,
\[ 
e^2 \,:=\, (\Lambda /v)^4
\,.
\label{eq:Se2}
\]
We will treat $e^2$ as a free parameter with the constraint that  $e^2 \gg 1$ to guarantee the applicability of our EFT construction.

To summarise our discussion so far, we have established that 
in the limit of the frozen out Higgs field and the decoupled gauge field, 
\[
s(x)\to v/\sqrt{2}\,, \quad W_\mu\to 0\,,
\label{eq:frozen}
\]
our electro-weak Lagrangian with the dimension-8 EFT operator \eqref{eq:stab8} reduces to the Lagrangian of the Skyrme model~\cite{Skyrme:1961vq},
\[
 {\cal L}_{\rm Skyrme}\,=\, 
 \frac{v^2}{4} \, {\rm tr} \left(\partial^\mu  U^\dagger \partial_\mu U\right)\,+\,
\frac{1}{32\, e^2}\, {\rm tr} \left[U^\dagger \partial_\mu U\,,\,  U^\dagger \partial_\nu U \right]^2 \;.
\label{eq:SkyrmeL}
\]
This model has topological solitons -- the skyrmions. These are classical solutions for the $U(x)$ field configurations,
with the Higgs and gauge fields being frozen out as dictated by \eqref{eq:frozen}. A single skyrmion has the winding number $n=1$,
an anti-skyrmion has $n=-1$ and the perturbative vacuum with no skyrmions corresponds to $n=0$. Due to their topological stability,
skyrmions with different charges $n \in {\mathbb Z}$ are separated by infinite energy barriers from one another and from the perturbative
vacuum. 
Skyrmion solitons of the model~\eqref{eq:SkyrmeL} are found using the the Skyrme ansatz, i.e. the so-called hedgehog configuration~\cite{Skyrme:1961vq}
\[
U_{\rm Sk}^{(0)} ({\bf x})\,=\, \exp[ i \tau^a \hat{x}^a F(r)]\,,
\label{eq:hedgeh}
\]
where $\tau^a$ are the Pauli matrices, $\hat{x}^a$ is the unit radius-vector and $r=|{\bf x}|$. The skyrmion function $F(r)$ has the boundary conditions $F(r) \to n\pi$ at $r\to 0$ and $F(r) \to 0$ at $r\to \infty$, and is found numerically.  $n \in {\mathbb Z}$ is the skyrmion number, and it can also be computed by evaluating the integral on the {\it r.h.s.} of~\eqref{eq:nHdef} on \eqref{eq:hedgeh}.
The mass and size of a single skyrmion ($n=1$)
are known~\cite{Adkins:1983ya} and in our notation given by,   
\[
M_{\rm Sk}^{(0)} \,\simeq\, 73\, \frac{v}{e} \,=\, 73 \,\frac{v^3}{\Lambda^2}\,,
\qquad {\rm and} \qquad
R_{\rm Sk}^{(0)} \,\simeq\,  \frac{2}{ve}\,=\, \frac{2v}{\Lambda^2}
\,.
\label{eq:MR}
\]
The superscript$^{(0)}$ indicates that the idealized settings \eqref{eq:frozen} were applied in computing these quantities using the classical
solution of the model \eqref{eq:SkyrmeL} (and the quantum corrections arising from skyrmion quantisation were also neglected).

Historically, in particle physics, skyrmions were successfully identified with baryons, specifically nucleons, emerging as topological solitons in the non-linear sigma model description
of strong interactions~\cite{Skyrme:1961vq,Witten:1979kh,Witten:1983tx,Adkins:1983ya,Zahed:1986qz}, 
\[
 {\cal L}_{\rm\,  Skyrme}^{\rm\, \,\,strong}\,=\, 
 \frac{F_\pi^2}{16} \, {\rm tr} \left(\partial^\mu  U^\dagger \partial_\mu U\right)\,+\,
\frac{1}{32\, e^2}\, {\rm tr} \left[U^\dagger \partial_\mu U\,,\,  U^\dagger \partial_\nu U \right]^2 \;,
\label{eq:SkyrmeLst}
\]
where $U(x)$ are the meson (pion) fields, and $F_{\pi}$ is the pion decay constant. 
The single skyrmion is a topological soliton of \eqref{eq:SkyrmeLst} that carries  one unit of the topological charge~\eqref{eq:nHdef}.
The latter is identified with the baryon number or the nucleon charge~\cite{Witten:1983tx} in the model \eqref{eq:SkyrmeLst}.

\medskip
In our case skyrmions arise instead in the electroweak sector of the SM in the EFT description that includes the dimension-8 operator~\eqref{eq:stab8}, which in the idealised decoupling limit resulted in the model~\eqref{eq:SkyrmeL}.

\medskip
We now return to the general case described by the Lagrangian (no longer assuming the decoupling limit),
\begin{eqnarray}
{\cal L} &=& -\frac{1}{2g^2}\,{\rm tr} \, W^{\mu \nu}W_{\mu\nu} \,+\,
\frac{1}{2} {\rm tr} \left(\left(D^\mu  \Phi\right)^\dagger D_\mu \Phi\right)\,-\,
\frac{\lambda}{4} \left({\rm tr} \, \left(\Phi^\dagger \Phi\right)-v^2\right)^2\nonumber\\
&& +\,\frac{1}{8\Lambda^4}\,
 {\rm tr} \left((D_\mu\Phi)^\dagger D_\nu \Phi\,-\, (D_\nu\Phi)^\dagger D_\mu \Phi\right)^2.
\label{eq:LsmSk}
\end{eqnarray}
On general grounds we expect that there are skyrmion solutions in this model but the skyrmions are not topological solitons.
Both, the presence of the gauge fields $W_\mu$ and the presence of the dynamical  Higgs field $s(x)$ (or $h(x)$), lift the topological protection of  skyrmions. The Higgs winding number \eqref{eq:nHdef}
is no longer a conserved quantity, as the skyrmion configuration with a non-vanishing $n_H$
can unwind itself via interactions with gauge and Higgs fields, for example when the Higgs field $s(x)=0$. The energy barrier separating the non-topological skyrmion from the perturbative vacuum is finite. Electroweak skyrmions can in principle be produced and they can decay as well in interactions involving vector and Higgs bosons.

Ambjorn  and Rubakov \cite{Ambjorn:1984bb} considered the effect of gauge interactions on the skyrmions in the model with a non-dynamical Higgs field. Ref.~\cite{Ambjorn:1984bb} 
showed that interactions with gauge fields destroys the skyrmion 
if the parameter $\xi =4e^2/g^2$ is less than a critical value $\xi_* \simeq 10.35$. 
On then other hand, for the values of $4e^2/g^2$ greater than 10.35, the skyrmion exists as a local minimum and it is separated by a finite
barrier from the trivial perturbative vacuum along the gauge field direction in the configuration space.
This translates into the gauge stability of the skyrmion requirement~\cite{Ambjorn:1984bb}, 
\begin{equation}
    M_{\text{Sk}} \lesssim  6.0 \frac{m_W}{\alpha_w} \sim \SI{15}{TeV}.
    \label{eq:gaugebound}
\end{equation}
Skyrmions with higher masses are unstable with respect to decays into gauge bosoins.

\medskip
Here, for the first time, we include the effects of the dynamical Higgs field, and will provide the description of electroweak skyrmions including the gauge-field and Higgs-field interactions in the context of the electroweak theory \eqref{eq:LsmSk}.
We will find that stability condition for the electroweak skyrmion against rapid decays into dynamical Higgs and gauge bosons becomes 
({\it cf.} eq.~\eqref{eq:Hgaugebound}),
\begin{equation}
    M_{\text{Sk}} \lesssim  \SI{10}{TeV},
    \label{eq:gaugebound}
\end{equation}
and, as such, is not dramatically lowered by the inclusion of physical Higgs fields, thus allowing for the existence of meta-stable and exponentially long-lived electroweak skyrmions with masses below 10 TeV.

\medskip
\subsection{Comment on the uniqueness of  the Skyrme term in EFT}
\medskip

One can ask what was special about selecting the dimension-8 operator in \eqref{eq:stab8} from a multitude of other possible choices available in the EFT. The main point here is that we require our non-topological soliton to become topologically protected in the limit where we have decoupled the gauge and the singlet Higgs fields.\footnote{While, a priori, one cannot exclude an occurrence of metastable minima of the energy that are not related to any symmetry or topology arguments in any limit, such hypothetical configurations would be accidental in the sense that they have no reason to exist and are not what we call  skyrmions here.}
In this case we have to be able to find stable minima of the action for the theory described by the  $U(x)$ field. Since the $U$-field is dimensionless, the general EFT description is the derivative expansion. The term with two derivatives is the kinetic term 
in \eqref{eq:SkyrmeL} that is already canonically normalised. The next term in the derivative expansion has to be a term with four derivatives. Witten argues in \cite{Witten:1983tx} that the Skyrme term is the unique four-derivative term that leads to a positive-definite Hamiltonian, and as such, is suitable for stabilising the energy functional of the skyrmion. Our expression \eqref{eq:stab8} is the lowest-dimensional embedding of the Skyrme term into the full EFT with dynamical gauge and singlet Higgs fields.

There exist higher-dimensional EFT operators that can stabilise skyrmions, but they would give subleading corrections to \eqref{eq:stab8}. To find them, one can either search for the suitable six-derivative operators, or alternatively find a higher-dimensional embedding of the four-derivative Skyrme term  \eqref{eq:stab8}.
An instructive example is provided by the dimension-12 operator,
\medskip
\[
{\cal L}_{\rm \, S12}\,=\, 
\frac{1}{2\Lambda^8}\,
 {\rm tr} \left[\Phi^\dagger D_\mu \Phi\,,\,  \Phi^\dagger D_\nu \Phi \right]^2,
\label{eq:stab12}
\]
which is an obvious direct embedding of \eqref{eq:stab8} in the full theory.
This operator, however is suppressed by 4 extra powers of the high scale $\Lambda$ relative to our dimension-8 operator~\eqref{eq:stab8}.

\medskip
\subsection{The vacuum structure, skyrmions and $(B+L)$ non-conservation}
\label{sec:2.4}
\medskip

To find skyrmion solutions we should search for local minima of the energy $E$
over all static configurations in the field space of the model
\eqref{eq:LsmSk}. This is the approach we will follow in the next section, using
numerical methods. Here we would like to first outline the general expected
structure of the energy landscape in our theory.

We consider those fields configurations that minimize the energy for fixed $n_H$ and $n_{\text{CS}}$. Neither $n_H$ nor $n_{\text{CS}}$,
defined in \eqref{eq:nHdef}-\eqref{eq:nCSdef},
 are gauge-invariant. They both change by an integer number under large gauge transformations, as $n_H \to n_H + N$ and $n_{\text{CS}} \to n_{\text{CS}} + N$. Small gauge transformations cannot change either of them. Thus, it is the difference between the two,
\begin{equation}
  n_{\text{Sk}} = n_H - n_{\text{CS}} \;,
  \label{eq:nSkdef}
\end{equation}
that is invariant under (large and small) gauge transformations. 
We will call the quantity $n_{\text{Sk}}$ on the right hand side of \eqref{eq:nSkdef}
the \emph{skyrmion number}, since the 1-skyrmion configuration has $n_{\text{Sk}} \simeq 1$ (or more precisely, $n_{\text{Sk}} = 1$ in
the limit where the skyrmion is a topological soliton, decoupled from the gauge and neutral Higgs fields).

There are several distinguished points in the $(n_H, n_{\text{CS}})$ field space. The trivial vacuum sits at $n_H = n_{\text{CS}} = 0$ and is known as the perturbative vacuum. The well-known gauge-transformed versions of it can be found at integer $n_H = n_{\text{CS}}\in {\mathbb Z}$ and are referred to as the large-pure-gauge vacua (all with the same degenerate vacuum energies). The unstable configuration at the mid point between any two neighbouring pure-gauge vacua is the electroweak sphaleron~\cite{Manton:1983nd}. Moving away from $n_{\text{Sk}} = 0$, one finds local minima of the energy functional: the skyrmion at $n_{\text{Sk}} \simeq 1$, the anti-skyrmion at $n_{\text{Sk}} \simeq -1$, and multi-skyrmion configurations at higher values of $|n_{\text{Sk}}|$. 

\begin{figure}
  \centering
  \includegraphics[width=0.45\textwidth]{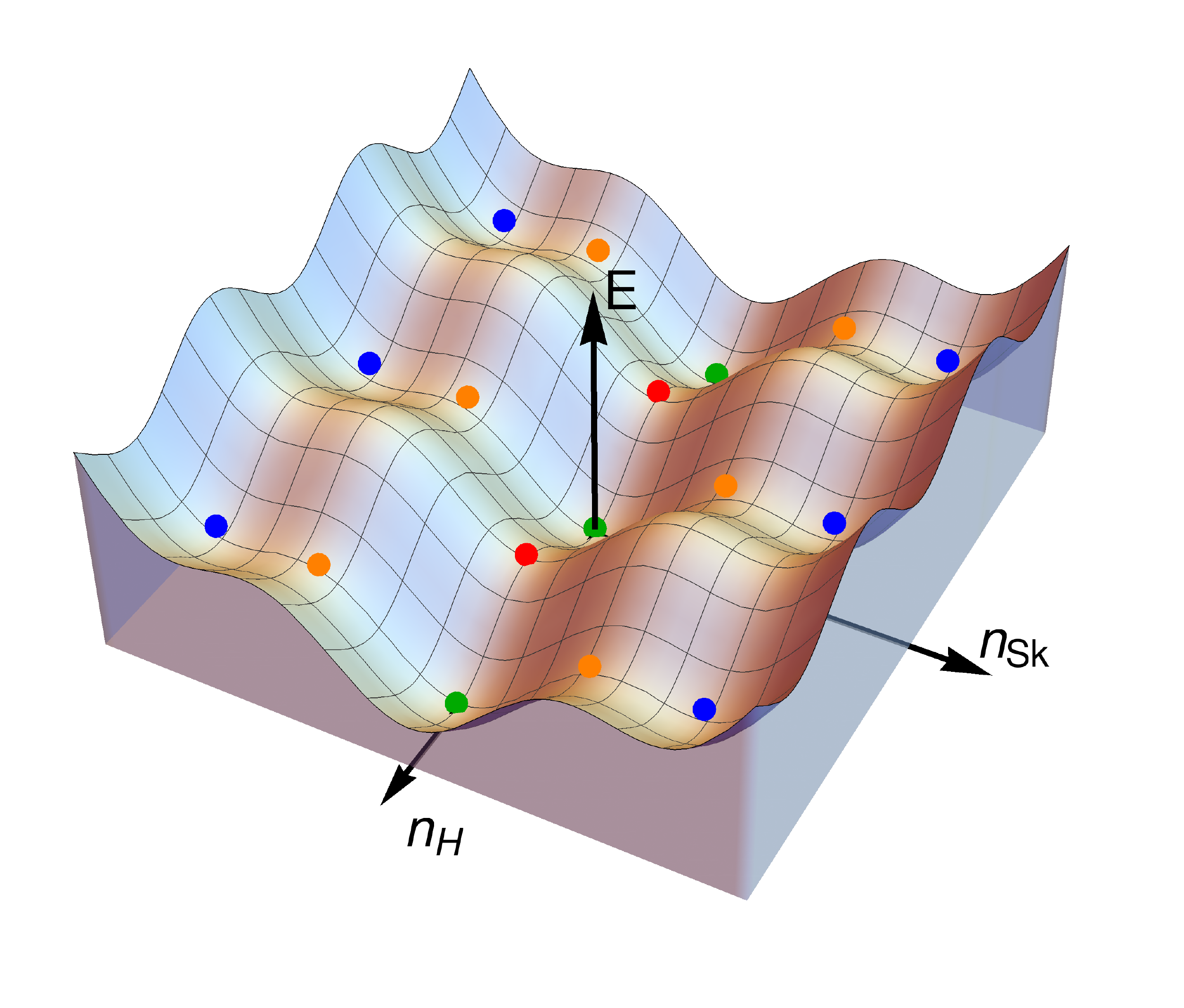}
  \includegraphics[width=0.45\textwidth]{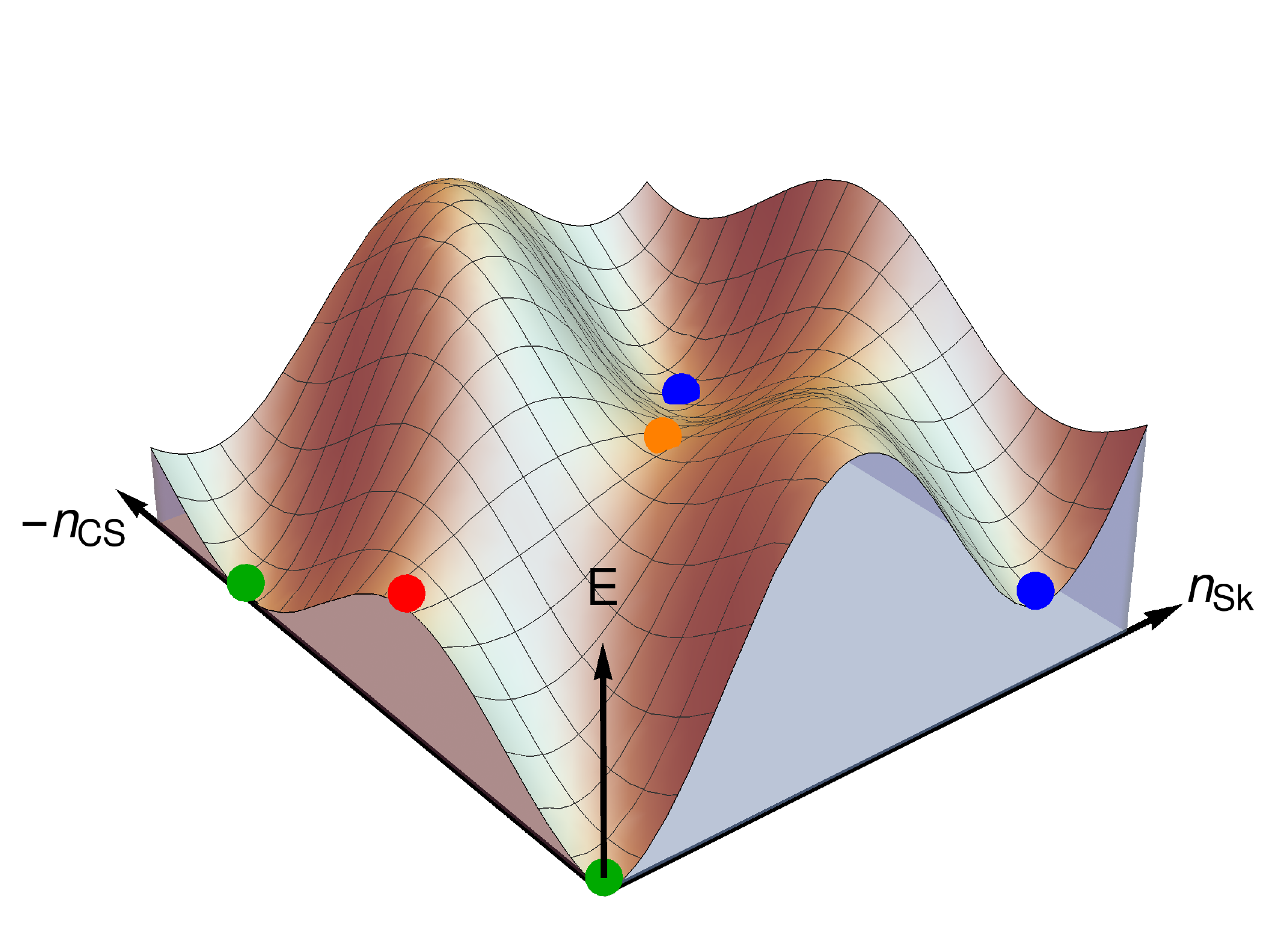}
  \caption{The energy profile of finite-energy field configurations parameterised by the $(n_{H},n_{\text{Sk}})$ coordinates in the plot on the left figure and by $(n_{\text{CS}},n_{\text{Sk}})$ in the figure on the right.
  Green dots are the pure-gauge vacuua with $n_{H}=n_{\text{CS}}\in \mathbb{Z}$ and $n_{\text{Sk}}=0$, red dots are the sphalerons, blue dots are the skyrmion and anti-skyrmion configurations $n_{\text{Sk}}=\pm1$ along with their large gauge transformations. Orange dots indicate the barriers at $n_{\text{Sk}} = \pm 1/2$ between the vacua and the (anti-)skyrmions. The energy axis $E$ resides at the origin, $n_H=n_{\text{Sk}}=n_{\text{CS}}=0$.}
  \label{fig:potential}
\end{figure}

A schematic picture of the energy functional and some of the distinguished points is shown in figure~\ref{fig:potential}. 
In  the plot on the left panel of figure~\ref{fig:potential} we use $n_H$ and $n_{\text{Sk}}$ to parameterise the field space, so the large-pure-gauge vacua (shown in green) are located at $n_{\text{Sk}}=0$ and integer $n_H$, with the sphalerons (shown in red) being at half-integer values of $n_H$
and $n_{\text{Sk}}=0$.\footnote{Indeed, it is easy to see that both, the pure-gauge vacua $n_H=n_{\text{CS}} \in {\mathbb Z}$ and the sphaleron at $n_H=n_{\text{CS}} =1/2$ have vanishing skyrmion number $n_{\text{Sk}}=0$.}
The skyrmion and anti-skyrmion configurations (shown as blue dots) correspond to $n_{\text{Sk}}=\pm 1$ and there is a new sphaleron-like saddle-points (in orange) separating the perturbative from the (anti)-skyrmion vacuum sectors.

The picture on the right panel of figure~\ref{fig:potential} depicts the same energy profile over the field configuration space with an alternative parameterisation, now in terms of $(n_{\text{CS}}, n_{\text{Sk}})$. In these coordinates, the pure-gauge vacua are at $(0,0), \,(\pm1,0),\,  \ldots\,$, the single-skyrmion vacua are at $(0,1), \,(\pm1,1),\,  \ldots\,$, the usual sphaleron is at $(\pm 1/2,0)$ and the new sphaleron-like barrier is at $(\pm1/2,1/2)$. It follows from this picture that the minimal finite energy path between the perturbative vacuum $(0,0)$ and the skyrmion $(-1,1)$ passes through the  saddle point  $(-1/2,1/2)$. We will compute this finite-energy trajectory numerically in the following section, see Fig~\ref{fig:energy-vs-n}.

As $\Lambda$ increases, the energy of the static skyrmion configuration (the skyrmion mass) approaches zero. When the Skyrme term is not present, the skyrmion becomes a texture, a rapidly shrinking configuration that decays into one of the pure-gauge vacua with $n_{\text{Sk}}=0$ through the (un)winding of either $n_H$ or $n_{\text{CS}}$~\cite{Turok:1990zg,vanderMeulen:2005sp}.

\medskip

We can use the gauge freedom \eqref{eq:gaugedef}
to select the unitary gauge, ${\cal U}_L(x) = U(x)^\dagger$, so that the Higgs field $\Phi(x)$ in \eqref{eq:Udef} is given just by the singlet field $s(x)$.
Static field configurations in the unitary gauge are:
\[ 
  W_0(\textbf{x}) \,=\, 0\,, \qquad
  W_i ({\bf x})\,, \qquad
  \Phi(\textbf{x}) \,=\, s({\bf x}) 1_{2 \times 2}\,.
\label{eq:stat_uni}
\]
If $s(\mathbf{x}) > 0$ for all $\mathbf{x}$, then $n_H = 0$, and thus, the skyrmion, having $n_{\text{Sk}} = 1$, is found for $n_{\text{CS}} = -1$.
The trivial vacuum is at $n_{\text{CS}} =0$ and $n_{\text{Sk}} =0.$

Transitions over or under the barriers that separate minima with different $n_{\text{CS}}$ will necessarily lead to baryon plus lepton number ($B + L$) violation, since the $B + L$ current in the Standard Model satisfies the anomalous Ward identity,
\begin{equation}
  \partial_\mu J^\mu_{B + L}
  =
  \frac{3}{8 \pi^2} \tr{W_{\mu\nu} \tilde{W}^{\mu\nu}}
  =
  6 \, \partial_\mu J^\mu_{\text{CS}},
\end{equation}
where $J^\mu_{\text{CS}} = \epsilon^{\mu\nu\rho\sigma} \tr{\left[W_\nu W_{\rho\sigma} +i (2/3) W_\nu W_\rho W_\sigma\right]} / 16\pi^2$ is the Chern-Simons current. The charge associated with the Chern-Simons current is just the Chern-Simons number $n_{\text{CS}} = \int d^3x J^0_{\text{CS}}$, and so, a change $\Delta n_{\text{CS}}$ in the Chern-Simons number implies a change in $B + L$,
\begin{equation}
  \Delta (B + L) = 6 \, \Delta n_{\text{CS}}.  
\end{equation}
For all SM fermions that are lighter than the skyrmion mass scale, a skyrmion production (or a skyrmion decay) in our theory,
\begin{equation}
(n_{\text{CS}}, n_{\text{Sk}})_{\rm space}:\qquad (0,0) \,\leftrightarrow (-1,1)\,,
\end{equation}
 will be accompanied by the fermion number for each fermion changing by one unit. When the skyrmion disappears, one net anti-fermion of each species will be produced, and if the skyrmion is produced, it will be accompanied by the newly minted SM fermion for each species.

If, on other hand, a given SM fermion (e.g. a top quark), is heavy relative to the skyrmion, no net change in the number of these fermions will occur. Instead the light skyrmion will itself carry the fermion number charge of the heavy fermion~\cite{Farhi:1995aq}. 
The criterium for distinguishing between light and heavy fermions is
\[
m_f R_{\text{Sk}} \ll 1\,,\quad {\rm or} \quad   m_f R_{\text{Sk}}\gg 1\,,
\]
where $m_f$ is the fermion mass and $R_{\text{Sk}}$ is the skyrmion size.

\medskip
\section{The skyrmion field and the energy profile in  configuration space}
\label{sec:sum}
\medskip

We now switch to the conventional complex doublet notation for the Higgs field \eqref{eq:phi_norm} and 
 write down the Lagrangian of our theory \eqref{eq:LsmSk} in the form,
 \begin{align}
  \mathcal{L}
  &=
    -\frac{1}{2g^2} \tr W_{\mu\nu} W^{\mu\nu}
    + |D_\mu \phi|^2
    -\lambda \( \phi^\dagger \phi - \frac{v^2}{2}\)^2
    + \frac{1}{\Lambda^2} \OSk \;,
\end{align}
where
\begin{align}
  \OSk = (D_{(\mu} \phi^\dagger D_{\nu)} \phi)^2 - |D_\mu \phi|^2.
\end{align}
We have used parentheses to denote symmetrisation of indices as $2t_{(\mu\nu)} \equiv t_{\mu\nu} + t_{\nu\mu}$. The energy for a static configuration is
\begin{align}
  E
  &=
    \int d^3x
    \Bigg\{
    \frac{1}{2g^2} \tr W_{ij} W^{ij}
    + |D_i \phi|^2
    + \frac{v^2 m_h^2}{8} \(\frac{2}{v^2} \phi^\dagger \phi - 1\)^2
  \nonumber \\
  &\phantom{= \int d^3x \Big\{}
    - \frac{1}{\Lambda^4} \((D_{(i} \phi^\dagger D_{j)} \phi)^2 - |D_i \phi|^4\)
    \Bigg\}.
    \label{eq:Lgauge-Sk_u}
\end{align}
We work in the unitary gauge, in which we use the parametrisation 
\begin{equation}
  \phi(x) = \frac{v \sigma(x)}{\sqrt{2}}
  \begin{pmatrix}
    0 \\ 1
  \end{pmatrix} \;,
\end{equation}
so that $\sigma(x)$ is a dimensionless field. In implementing the search for single-skyrmion (or anti-skyrmion) configurations, along with the with the finite-energy trajectories in the field configuration space connecting them to the perturbative vacuum, we impose the spherical ansatz~\cite{Witten:1976ck},
and write $\sigma = \sigma(r)$ and 
\begin{equation}
  W_i 
  =
  \frac{\Lambda^2}{v} \tau_a \(
    \epsilon_{ija} n_j \frac{f_1(r)}{r}
    + (\delta_{ia} - n_i n_a) \frac{f_2(r)}{r}
    + n_i n_a \frac{b(r)}{r}
    \),
\end{equation}
where $r^2 = (\Lambda^2/v)^2 \sum_i x_i^2$, $n_i = (\Lambda^2/v) x_i/r$, and $\tau_i$ are the Pauli matrices. In terms of these variables the energy is
\begin{equation}
  E = \frac{4 \pi v^3}{\Lambda^2} E_{\text{nat}},
\end{equation}
where the dimensionless quantity  $E_{\text{nat}}$ is the energy functional in natural units,
\begin{align}
  E_{\text{nat}} &= \int_0^\infty dr
      \Bigg\{
      \frac{\Lambda^4}{v^2 m_W^2} \left[
      \left(f_1' - 2f_2\frac{b}{r}\right)^2
      + \left(f_2' - (2f_1 -1)\frac{b}{r}\right)^2
      + \frac{2}{r^2} (f_1^2 + f_2^2 - f_1)^2
      \right]
      \nonumber \\
    &\phantom{= \int dr\Bigg\{+}
      + \frac{r^2}{2} (\sigma')^2
      + \sigma^2 \(f_1^2 + f_2^2 + \frac{b^2}{2}\)
      + \frac{m_h^2 v^2 r^2}{8 \Lambda^4} (\sigma^2 - 1)^2
      \nonumber \\
    &\phantom{= \int dr\Bigg\{+}
      + (f_1^2 + f_2^2)
      \left[
      (\sigma')^2 + \frac{\sigma^2}{r^2} \(b^2 + \frac{f_1^2 + f_2^2}{2}\)
      \right]
  \Bigg\}.
  \label{eq:energy}
\end{align}

In order to explore the energy functional in the direction of increasing $n_{\text{Sk}}$ we define the following new coordinate in field space~\cite{Ambjorn:1984bb},
\begin{equation}
  n_W = \frac{1}{24\pi^2} \int d^3x \, \epsilon_{ijk}\,  {\rm tr} \left(iW_i W_j W_k \right) \\
  = \frac{2}{\pi} \int_0^\infty dr \, \frac{b}{r} \, (f_1^2 + f_2^2),
  \label{eq:cs-number}
\end{equation}
where in the second equality we have used the spherical ansatz. This is a simplified version of the Chern-Simons number: we have $n_W = n_{\text{CS}}$ when $W_i$ is a pure gauge. So they both assign the same integer label to each pure-gauge configuration, but give different interpolations between them.

Concerning the boundary conditions for the $f_1$, $f_2$, $b$ and $\sigma$ functions, regularity of $\Phi$ and $W_i$ at $r = 0$ requires that
\begin{equation}
  f_1(0) = f_1'(0) = f_2(0) = b(0) = f_2'(0) - b'(0) = \sigma'(0) = 0.
\end{equation}
At large $r$, we impose
\begin{equation}
  f_1(r) \propto f_2(r) \propto b(r) \propto e^{-r}, \qquad \sigma - 1 \propto e^{-r m_h/ m_W},
\end{equation}
which is sufficient to ensure that the boundary conditions \eqref{eq:bc-topological} are satisfied.

\begin{figure}
  \centering
  \includegraphics[width=0.45\textwidth]{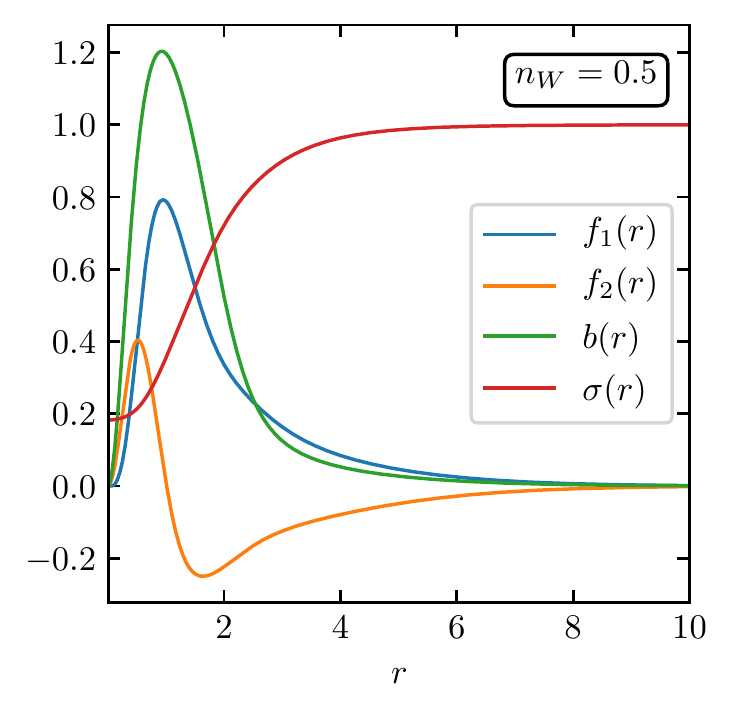}
  \includegraphics[width=0.45\textwidth]{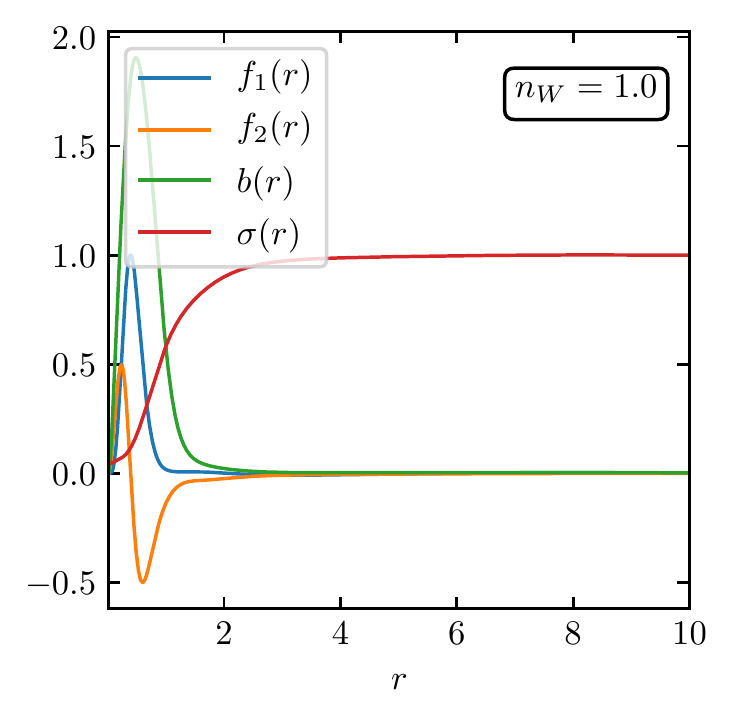}
  \caption{Field configurations for $n_W = 1/2$ and $n_W = 1$ at fixed $\Lambda = \SI{200}{GeV}$ as functions of $r$.}
  \label{fig:field-configuration}
\end{figure}

\begin{figure}
  \centering
  \includegraphics[width=0.45\textwidth]{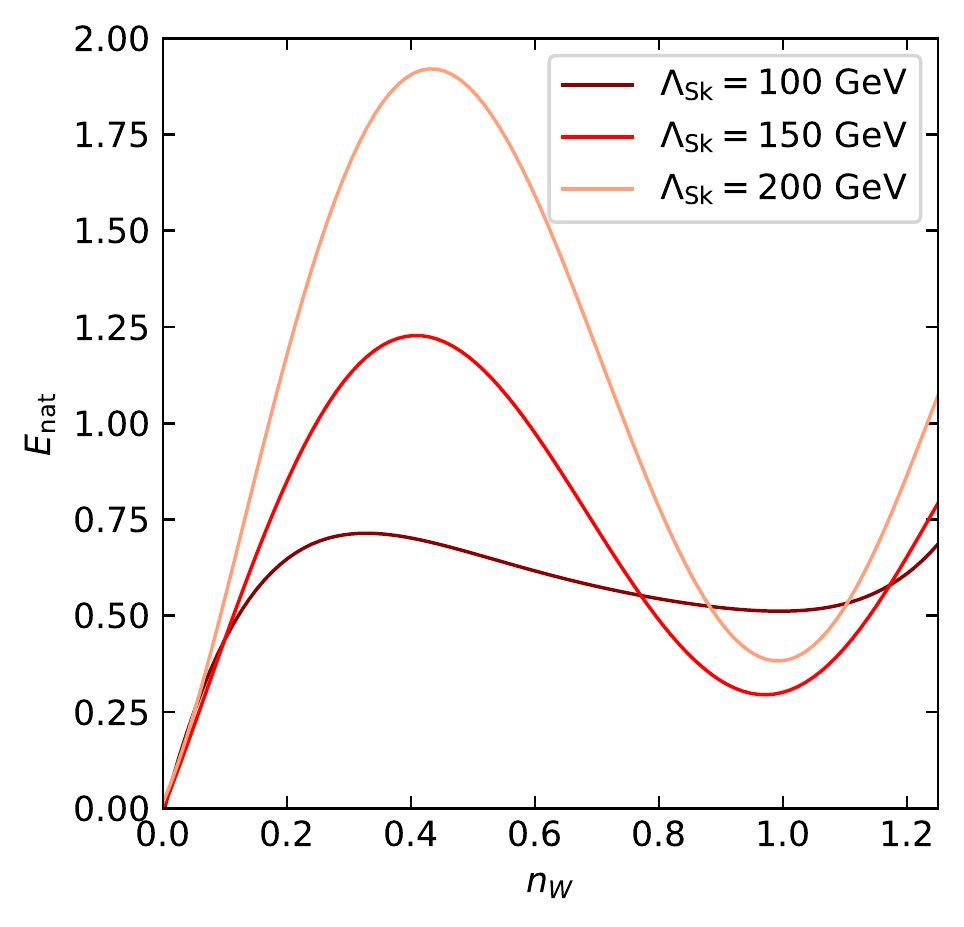}
  \includegraphics[width=0.45\textwidth]{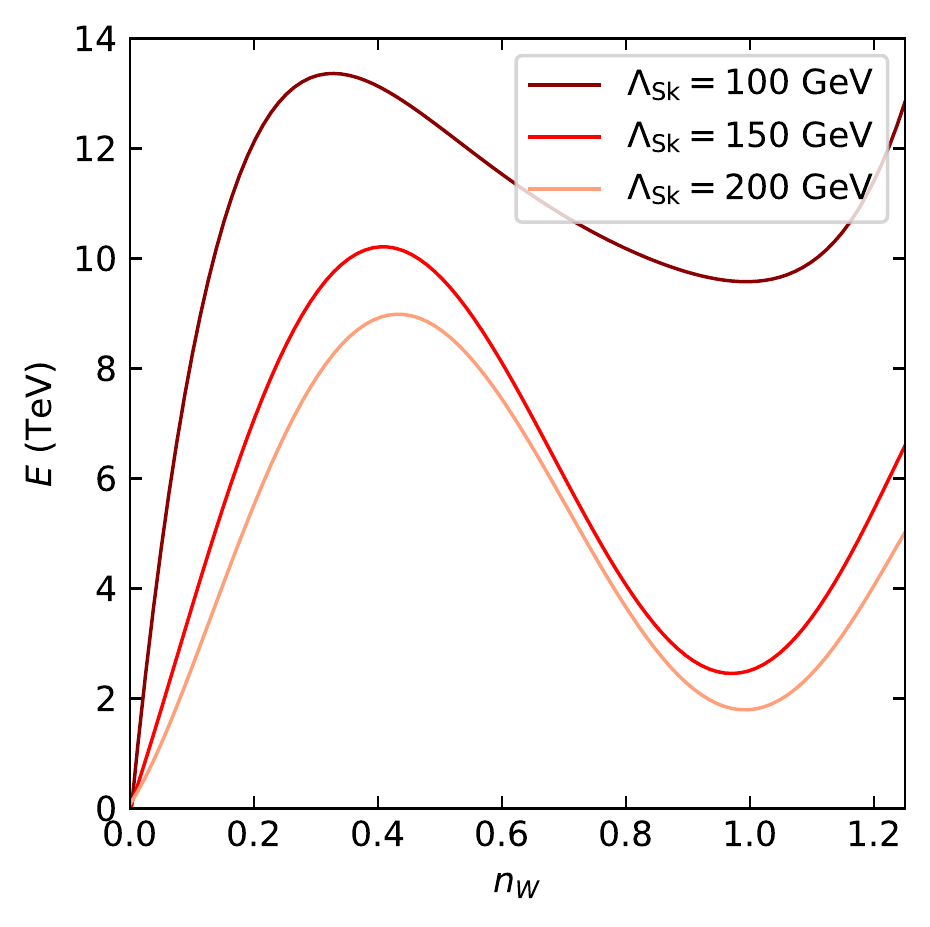}
  \caption{Energy as a function of $n_W$ for different values of $\Lambda$. The figure on the left shows the dimensionless quantity $E_{\text{nat}}$ and the figure on the right gives the corresponding physical $E$ in TeV. The (anti)-skyrmion solution corresponds to the local minimum in the vicinity of $n_W\simeq 1$ and the new spaleron-like barrier is around  $n_W\simeq 1/2$. Perturbative vacuum is the global minimum at $n_W=0.$}
  \label{fig:energy-vs-n}
\end{figure}

\begin{figure}
  \hspace{-20pt}
  \includegraphics[width=1.1\textwidth]{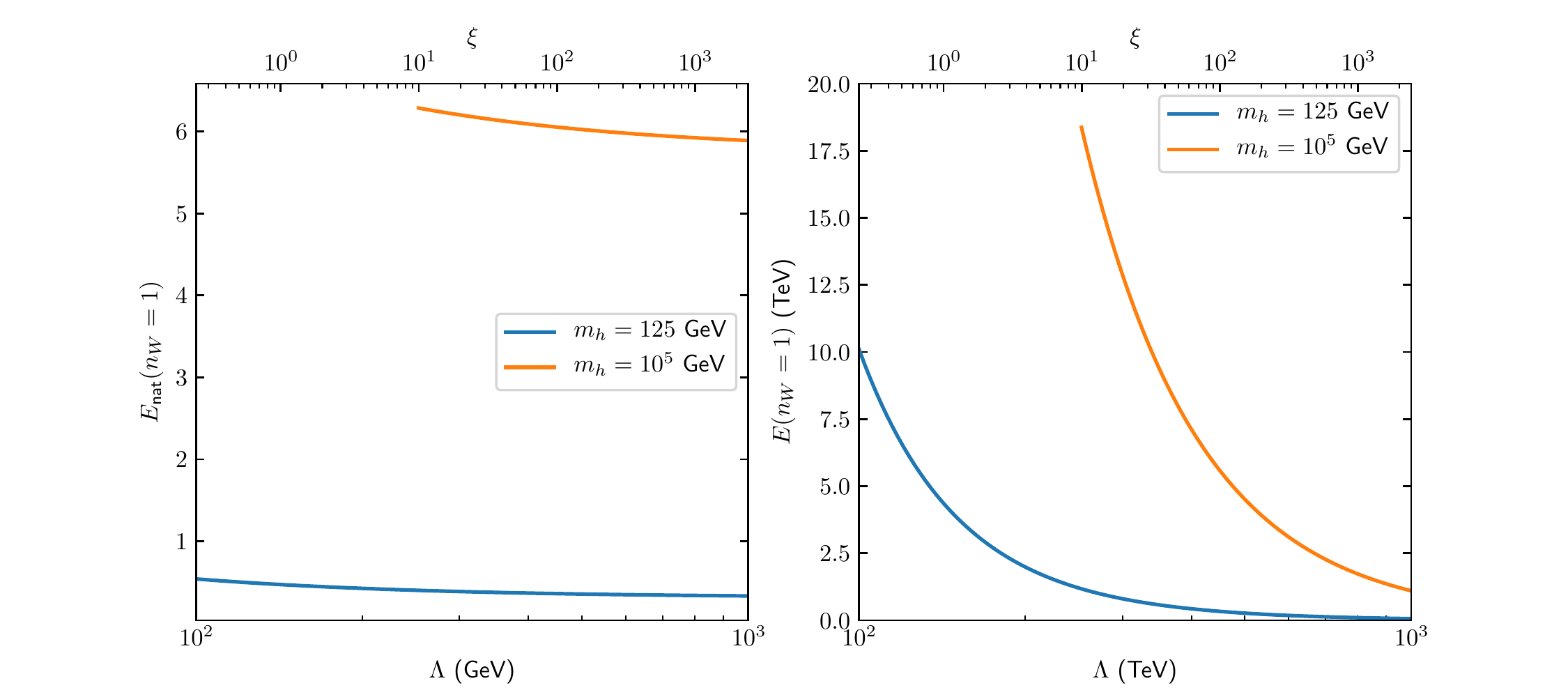}
  \caption{Energy in natural $E_{\text{nat}}$ and physical $E$ units, at $n_W = 1$, as a function of $\Lambda$ for $m_h = \SI{125}{GeV}$ and $m_h = \SI{e5}{GeV}$.}
  \label{fig:mass-vs-Lambda}
\end{figure}

Using the numerical method described in appendix~\ref{app:neural-nets}, we search for the minimal energy configuration with fixed values of $n_W$ and $\Lambda$. Some examples of the solutions we obtain are shown in figure~\ref{fig:field-configuration}. We find that, for $\Lambda \gtrsim \SI{100}{GeV}$, there is a minimum at $n_W \simeq 1$, the anti-skyrmion; and a barrier at $n_W \simeq 1/2$, separating the vacuum and the anti-skyrmion. This is shown in figure~\ref{fig:energy-vs-n} for several values of $\Lambda$. The mass $M_{\text{Sk}}$ of the skyrmion is approximately the energy $E$ at $n_W = 1$. In figure~\ref{fig:mass-vs-Lambda}, we show that the mass scales as $M_{\text{Sk}} \propto v^3 / \Lambda^2$ for large $\Lambda$. This is the same scaling as in eq.~\eqref{eq:MR}, and it is equivalent to $E_{\text{nat}}(n_W = 1)$ being approximately constant in $\Lambda$. We find that
\begin{equation}
  M_{\text{Sk}} \simeq E_{\text{nat}}(n_W = 1) \frac{4\pi v^3}{\Lambda^2} \simeq
  0.35 \frac{4\pi v^3}{\Lambda^2}.
\end{equation}
For low values of $\Lambda$, we get slight oscillation of $E_{\text{nat}}(n_W = 1)$ around its limiting value, with it being higher than the limiting value for $\Lambda \simeq \SI{100}{GeV}$ and lower for $\Lambda \simeq \SI{200}{GeV}$. We also check the value of $M_{\text{Sk}}$ in the decoupling limit by taking the Higgs mass to be very large ($m_h = \SI{e5}{GeV}$). We get $E_{\text{nat}}(n_W = 1)|_{\text{decoupling}} \simeq 6$, in agreement with ref.~\cite{Ambjorn:1984bb}. 

\medskip 

Below the critical value $\Lambda = \Lambda_{\text{crit}} \simeq \SI{100}{GeV}$ there is no local minimum at $n_W \simeq 1$ and only the perturbative vacuum remains. This implies an upper bound on the mass of the single skyrmion or single anti-skyrmion:
\begin{equation}
    M_{\text{Sk}} \lesssim E_{\text{nat}}(n_W = 1, \Lambda = \Lambda_{\text{crit}}) \frac{4\pi v^3}{\Lambda_{\text{crit}}^2} \simeq \SI{10}{TeV}.
    \label{eq:Hgaugebound}
\end{equation}

In order to compute the size $R_{\text{Sk}}$ of the skyrmion, we take the average $\left< r^2 \right>$ of the squared radius, with distribution given by the gauge-winding-number density as
\begin{align}
  R_{\text{Sk}}^2 = \left(\frac{v}{\Lambda^2}\right)^2 \left< r^2 \right>
  &=
    \left(\frac{v}{\Lambda^2}\right)^2 \frac{1}{24\pi^2} \int d^3x \, r^2 \, \epsilon_{ijk}\,  {\rm tr} \left(iW_i W_j W_k \right)
  \\
  &= \left(\frac{v}{\Lambda^2}\right)^2 \int dr \, r b (f_1^2 + f_2^2),
\end{align}
where in the last equality we have used the spherical ansatz. From our numerical solutions, we obtain
\begin{equation}
  R_{\text{Sk}} \simeq 0.6 \frac{v}{\Lambda^2}.
\end{equation}

It is worthwhile to note that that the spherical ansatz is consistent with the symmetry properties of single skyrmions, anti-skyrmions and the sphaleron-like saddle-point configurations at the top of the barrier. However, we should not expect to find energetically stable 
multi-skyrmion solutions using the form dictated by the spherical ansatz. The point is that multi-skyrmion solutions with $n_W \ge 2$ would be $O(3)$ symmetric only if the positions of the constituent single skyrmions were to coincide. Such configurations are known to be unstable and, in fact, they were shown in \cite{Brihaye:1989ej} to have higher energies than $n_W$  single skyrmions at infinite separations. Hence, in the above we concentrated on single (anti)-skyrmion configurations and the energy profile in figure~\ref{fig:energy-vs-n}  is the interval $0\le |n_W| \lesssim 1.2$.

\medskip
\section{UV completions of the electroweak Skyrme EFT}
\label{sec:ext}

Since the Skyrme term appears in the SMEFT as a non-renormalizable operator $\OSk$, one might wonder whether there are UV completions of the SMEFT that generate it. It turns out that it generically appears in simple renormalizable weakly-coupled extensions of the SMEFT with extra fields. An example, in which $\OSk$ is the only dimension-8 four-derivative operator that is generated, is a $SU(2)$ triplet vector boson $V$ with the following Lagrangian:
\begin{equation}
  \mathcal{L}_{\text{UV}} =
  \frac{1}{2} \left(
    D_\mu V^a_\nu D^\nu V^{a\mu}
    - D_\mu V^a_\nu D^\mu V^{a\nu}
    + M^2 V^a_\mu V^{a\mu}
    \right)
    + g_V V^a_\mu \; 2\im\left(\phi^\dagger \sigma^a D^\mu \phi\right).
  \label{eq:triplet}
\end{equation}
The mass term is written here explicitly, but it could be generated by the Higgs mechanism using an extra scalar. Integrating out $V$ gives:
\begin{align}
  \mathcal{L}_{\text{eff}}
  &=
    2 g^2_V
    \im(\phi^\dagger \sigma^a D_\mu \phi)
    \left[
    \frac{1}{M^2} \eta^{\mu\nu}
    + \frac{1}{M^4} \left(D^\nu D^\mu - D^2 \eta^{\mu\nu}\right)
    + O\left(\frac{1}{M^6}\right)
    \right]
    \im(\phi^\dagger \sigma^a D_\nu \phi)
    \label{eq:eff-lag-1}
  \\
  &=
    \frac{g_V^2}{2 M^2} [\im(\phi^\dagger \sigma^a D_\mu \phi)]^2
    + \frac{g_V^2}{M^4}
    \left(
    \OSk
    + 2 T
    \right)
    + O\left(\frac{1}{M^6}\right),
    \label{eq:eff-lag-2}
\end{align}
where $T$ is defined as
\begin{align}
  T
  &\equiv
    \epsilon^{abc} W^a_{\mu\nu} \im(\phi^\dagger \sigma^b \phi) \im(\phi^\dagger \sigma^c \phi)
    + \left[\im(\phi^\dagger \sigma^a D^2 \phi)\right]^2
    + \im(\phi^\dagger \sigma^a D^2 \phi) \tr(D_\mu \phi^\dagger \sigma^a D^\mu \phi)
    \nonumber \\
  &\phantom{=}
    + \im(\phi^\dagger \sigma^a D_\mu \phi) \im(D^\mu \phi^\dagger \sigma^a D^2 \phi)
    + \im(\phi^\dagger \sigma^a [D_\mu, D_\nu] \phi) \im(D^\mu \phi^\dagger \sigma^a D^\nu \phi)
    \nonumber \\
  &\phantom{=}
    + \tr(D_\mu \phi^\dagger \sigma^a [D_\mu, D_\nu] \phi) \tr(\phi^\dagger \sigma^a D^\nu \phi)
    + \tr([D_\mu, D_\nu] \phi^\dagger \sigma^a D^\mu \phi) \tr(D^\nu \phi^\dagger \sigma^a\phi)
    \nonumber \\
  &\phantom{=}
    + \tr(\phi^\dagger \sigma^a D^2 D_\nu \phi) \tr(\phi^\dagger \sigma^a D^\nu \phi).
\end{align}
All of the operators in $T$ can be eliminated in favor of others with less derivatives using field redefinitions (for those containing $D^2\Phi$) and the relation $[D_\mu, D_\nu] \sim F_{\mu\nu}$. The order-$1/M^2$ part of the effective Lagrangian is, in terms of Warsaw-basis operators~\cite{deBlas:2017xtg}:
\begin{align}
  \mathcal{L}_{\text{eff}}
  &=
    \frac{8 g_V^2 m_W^2 \kappa^2}{M^2} (\phi^\dagger \phi)^2
    - \frac{8 g_V^2 m_W^2 \kappa^2}{v^2 M^2} (\phi^\dagger \phi)^3
    - \frac{3 g_V^2}{2 M^2} (\phi^\dagger \phi) \square (\phi^\dagger \phi)
    \nonumber \\
  &\phantom{=}
    -  \frac{g_V^2}{M^2} \sum_\psi \left[
    y^*_\psi (\phi^\dagger \phi) (\bar{\psi}_L \phi \psi_R) + \text{h.c.}
    \right]
    + O\left(\frac{1}{M^4}\right).
\end{align}
The strongest bound comes from the $(\phi^\dagger \phi)\square(\phi^\dagger\phi)$ operator, because its coefficient is negative in this model, and the current limits almost rule out negative values. The 2$\sigma$-limit on this coefficient from ref.~\cite{Ellis:2018gqa} implies
\begin{equation}
  \frac{\Lambda}{\sqrt{g_V}} \simeq \frac{M}{g_V} \gtrsim \SI{1.2}{TeV} \;.
\end{equation}
Other UV models might generate the Skyrme term in a similar way, but extra 4-derivative order-$1/M^4$ operators are generated. For example, a scalar singlet with Lagrangian
\begin{equation}
  \mathcal{L}_{\text{UV}}
  =
  - \frac{1}{2} S (D^2 + M^2) S + \kappa_S S |\phi|^2,
\end{equation}
generates the effective Lagrangian
\begin{align}
  \mathcal{L}_{\text{eff}}
  &=
    \frac{\kappa_S^2}{2 M^2} |\phi|^4
    - \frac{\kappa_S^2}{2 M^4} |\phi|^2 \square |\phi|^2
  \\
  &\phantom{=}
    + \frac{\kappa_S^2}{2 M^6} \left(
    - \OSk
    + (D_{(\mu} \phi^\dagger D_{\nu)} \phi)^2
    + (\phi^\dagger D^2 \phi + \hc)^2
    +  (\phi^\dagger D^2 \phi + \hc) |D_\mu\phi|^2
    \right).
\end{align}

\begin{figure}
  \centering
  \includegraphics[width=0.4\textwidth]{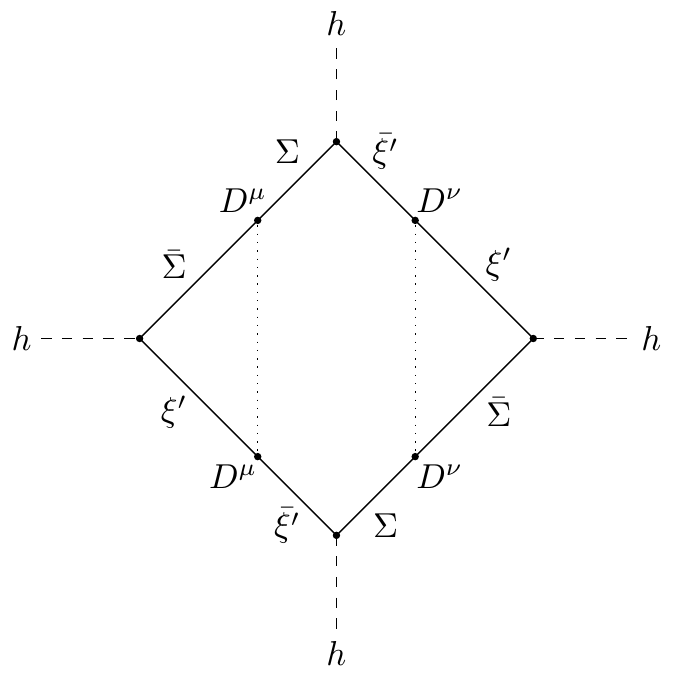}
  \includegraphics[width=0.4\textwidth]{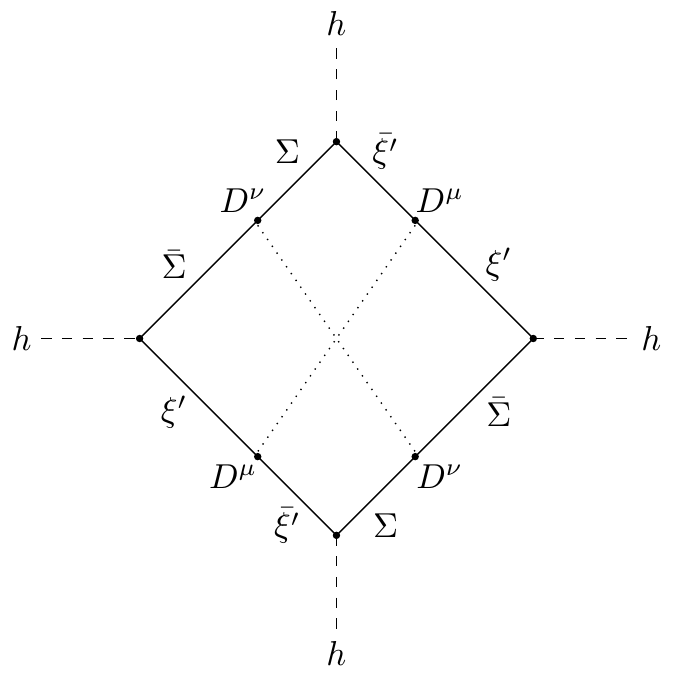}
  \caption{Diagramatic contribution to the Skyrme term before integrating out the heavy degrees of freedom $\Sigma$ and $\xi'$ defined in eqs.~(\ref{vll_deg}) and (\ref{eq:vlike}). $D$ refers to the covariant derivative.}
  \label{fig:vll}
\end{figure}

Not only bosons can give rise to the Skyrme term. Models with vector-like leptons, similar to the one proposed in \cite{DasBakshi:2020ejz}, can induce this term as well. Specifically, by extending the Standard Model by three heavy vector-like lepton multiplets 
\begin{equation}
  \label{vll_deg}
  \Sigma_{L,R}=\begin{pmatrix} \eta \\ \xi	\end{pmatrix}_{L,R} : (1,2,-1/2), \;\;
  \eta'_{L,R} : (1,1,0),\;\;
  \xi'_{L,R} : (1,1,-1),
\end{equation}
where the quantum numbers are depicted in $SU(3)_C\times SU(2)_L \times U(1)_Y$ convention.
The most-general gauge-invariant renormalizable Lagrangian with such vector-like leptons can be written as
\begin{eqnarray}
\label{eq:vlike}
\mathcal{L}_{\text{VLL}}  &=&  \bar{\Sigma} ( i \slashed{D}_{\Sigma} - m_{_{\Sigma}}) \Sigma + \bar{\eta'} ( i \slashed{D}_{\eta'} - m_{{\eta'}}) \eta' + \bar{\xi'} ( i \slashed{D}_{\xi'} - m_{{\xi'}}) \xi' \nonumber\\
	&& - \left\lbrace \bar{\Sigma} \tilde{\phi} ( Y_{\eta_{_L}} \mathbb{P}_L + Y_{\eta_{_R}} \mathbb{P}_{R}) \eta' + \bar{\Sigma} \phi ( Y_{\xi_{_L}} \mathbb{P}_{L} + Y_{\xi_{_R}} \mathbb{P}_{R}) \xi' + \text{h.c.}  \right\rbrace,
\end{eqnarray}
where, $Y_i$'s are the complex Yukawa couplings, 
$m_{_{\Sigma}}, m_{\eta'}$, and $m_{\xi'}$ are the masses of $\Sigma$, $\eta'$ and $\xi'$, respectively. $\mathbb{P}_{L}(\mathbb{P}_{R})$ are the left (right) chiral projection operator. The contribution to the Skyrme term from this UV model at 1-loop level can be captured pictorially in figure~\ref{fig:vll}. 

Thus, the Skyrme term can be induced rather generically by Standard Model extensions with particles of spins 0, 1/2 or 1. The presence of any of the particles introduced here, i.e. singlet scalar, vectorlike leptons or triplet vector bosons, would give rise to the Skryme term, respectively.

\section{Electroweak Skyrmion phenomenology}
\label{sec:pheno}

Following the discussion of Sec.~\ref{sec:2.4} the skyrmion production must be accompanied by $B+L$ violation in the same way as for the electroweak instanton / sphaleron transitions. If a skyrmion could be produced at collider experiments, it would provide a striking signature that could be easily separated from Standard Model backgrounds. Unfortunately, the direct production of electroweak skyrmions in a collider experiment is highly unlikely for the same reasons as inducing electroweak  transitions across the sphaleron barrier in 2 particle collisions is expected to be exponentially suppressed at any energies, below or above the sphaleron barrier~\cite{Banks:1990zb,Bezrukov:2003er}. Unsuppressed B+L violating processes in the Standard Model should have of order $1/\alpha_w \gg 1$ particles in the initial as well as the final states. This implies that such inter-vacua transitions for the vacua separated by the electro-weak sphaleron-size barriers will most likely be unobservable at future colliders at arbitraryly high energies, 
in agreement with the calculations in \cite{Bezrukov:2003er} and \cite{Khoze:2020paj}.

\medskip
\subsection{Probing the Skyrme term at collider experiments}
\medskip

\begin{figure}
  \centering
  \includegraphics[width=0.9\textwidth]{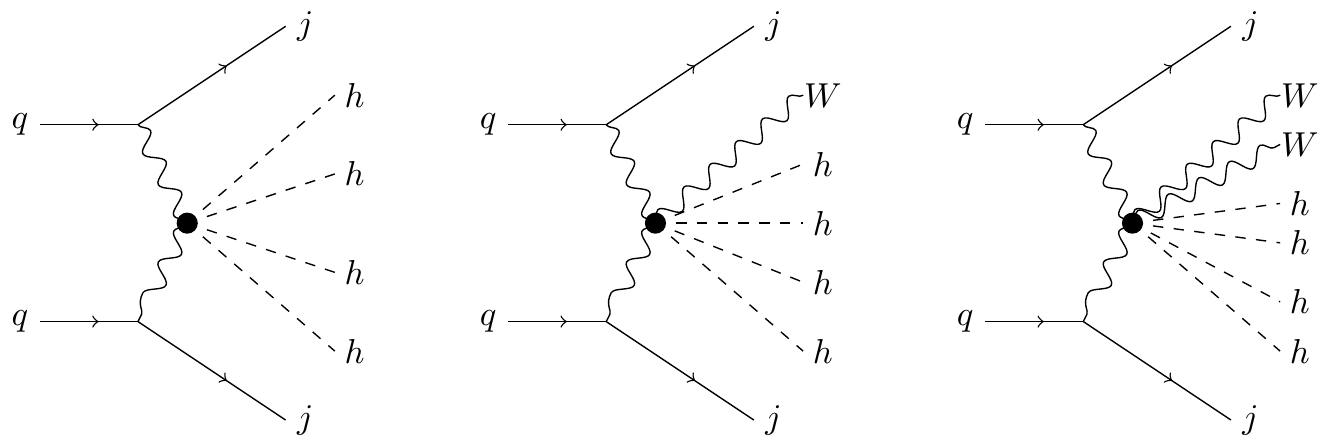}
  \caption{4-Higgs production in association with 0--2 W bosons, through the dimension-8 skyrmion operator $\OSk$, which is denoted by a dot in the diagrams.}
  \label{fig:vbf}
\end{figure}

The SMEFT dimension-8 realization \eqref{eq:stab8} of the Skyrme term can be probed at colliders through processes involving Higgs and electroweak gauge bosons. Both at hadron and lepton colliders, one can generate a pair of vector bosons that interact through $\OSk$, producing Higgs and/or vector bosons. The signal-to-background ratio is expected to be optimal for those processes with multiple Higgs bosons in the final state. The corresponding diagrams are shown in figure~\ref{fig:vbf}, for the case of a hadron collider. The weak-boson fusion (WBF) production cross sections for them can be parametrized as
\begin{equation}
  \sigma =
  A
  \left(\frac{\sqrt{s}}{\SI{14}{TeV}}\right)^B
  \left(\frac{\SI{1}{TeV}}{\Lambda}\right)^8,
\end{equation}
where $\sqrt{s}$ is the center-of-mass energy of the two beams. We simulate the $pp \to jj hhhh$ process using \texttt{MadGraph} for different values of $\sqrt{s}$ and $\Lambda$. We use the cuts $m_{jj} > \SI{400}{GeV}$, $p_{T,j} > \SI{30}{GeV}$, $|\eta_j| < 5$, $\Delta R_{jj} > 0.8$ and $|\eta_{j_1} - \eta_{j_2}
| > 2.5$, where $m_{jj}$ is the invariant mass of the two jets, $p_{T,j}$ and $\eta_j$ are the transverse momentum and rapidity of any of the two, and $\Delta R_{j_1 j_2}$ is the distance between them in the space of rapidity and azimuthal angle. To reconstruct reconstruct all the Higgs bosons, we apply b-tagging to each of the 6 b-jets, assuming a tagging efficiency of 80\%. We obtain
\begin{equation}
  A = \SI{2.70}{pb} \qquad \text{and} \qquad B = 8.93.
\end{equation}

For fixed $\sqrt{s}$, only low-enough values of $\Lambda$ will allow this process to be observable. We consider the channel in which three of the Higgs bosons decay into $b\bar{b}$ and one into $\gamma\gamma$, and require that 300 events are produced. We consider this a conservative estimate for the number of signal events to show a statistically significantly excess over Standard Model background events in this final state. Inclusive Di-Higgs production in $b\bar{b}\gamma\gamma$ shows sensitivity over backgrounds with a similar number of events \cite{Cepeda:2019klc,Contino:2016spe}. 
Thus, the process with WBF cuts for the tagging jets and four Higgs resonances should provide enough handles to control the backgrounds. 

A high-energy muon collider of $\sqrt{s}=14$ TeV has a sizeable WBF cross section. Due to its much cleaner environment and reduced QCD background, we only require 10 events for the discovery of Skyrme-term-induced processes. We, however, impose again 6 tagged b-jets.

We find limits on $\Lambda$ for various collider energies and luminosities:
\begin{align}
  \Lambda < \SI{58}{GeV}
  \qquad
  &\text{for } \sqrt{s} = \SI{14}{TeV}, \, \smallint dt \, L = \SI{300}{fb^{-1}},
  \\
  \Lambda < \SI{77}{GeV}
  \qquad
  &\text{for } \sqrt{s} = \SI{14}{TeV}, \, \smallint dt \, L = \SI{3000}{fb^{-1}},
  \\
  \Lambda < \SI{320}{GeV}
  \qquad
  &\text{for } \sqrt{s} = \SI{50}{TeV}, \, \smallint dt \, L = \SI{3000}{fb^{-1}},
  \\
  \Lambda < \SI{690}{GeV}
  \qquad
  &\text{for } \sqrt{s} = \SI{100}{TeV}, \, \smallint dt \, L = \SI{3000}{fb^{-1}}.
\end{align}
corresponding to LHC, HL-LHC and hh-FCC, respectively. For a 14 TeV muon collider in the process $\mu^+ \mu^- \to \nu_\mu \bar{\nu_\mu} hhhh $ we obtain
\begin{align}
  \Lambda < \SI{650}{GeV}
  \qquad
  &\text{for } \sqrt{s} = \SI{14}{TeV}, \, \smallint dt \, L = \SI{3000}{fb^{-1}}.
\end{align}
Consequently, in the final state with 6 b-quarks and two photons a higher-energy collider than the LHC is needed to probe values of $\Lambda$ that allow the formation of a skyrmion. If backgrounds can be confidently reduced in all-hadronic final states, see e.g. \cite{deLima:2014dta,Soper:2011cr,Soper:2014rya}, larger branching ratios can be exploited and higher scales $\Lambda$ surveyed.

\medskip
\subsection{Skyrmions as Dark Matter candidates}
\medskip

Although the skyrmion can unwind through electroweak instanton-like processes, such processes are highly suppressed, thereby rendering the lifetime of a freely propagating skyrmion likely to be longer than the lifetime of the Universe~\cite{Gillioz:2010mr}. Consequently, an electroweak skyrmion can constitute a dark matter candidate. In a freeze-out scenario, the skyrmion abundance is set to
\begin{equation}
  \Omega h^2 \simeq \frac{\SI{3e-27}{cm^3 s^{-1}}}{\left<\sigma_{\text{ann}} \mathrm{v}\right>},
\end{equation}
where $\left< \sigma_{\text{ann}} \mathrm{v} \right>$ is the thermally-averaged cross section for the annihilation of two skyrmions into SM particles. As an order-of-magnitude estimate for the annihilation cross section $\sigma_{\text{ann}}$ we just take the skyrmion area
\begin{equation}
  \sigma_{\text{ann}} \simeq \pi R_{\text{Sk}}^2.
  \label{eq:sigma-area}
\end{equation}
One can get an upper bound on $\Lambda$ by requiring that the skyrmion abundance is at most the measured value of the dark matter abundance $\Omega h^2 \simeq 0.1$. Taking the velocity of the skyrmions at the freeze-out temperature to be $\mathrm{v} = 1/2$, we get $\Lambda \lesssim \SI{2}{TeV}$. This bound would be saturated if all of the dark matter was made of skyrmions. Although it depends on the rough approximation \eqref{eq:sigma-area}, it is relatively stable against corrections to it, since it is proportional to $\sigma_{\text{ann}}^{-1/4}$. Allowing $\sigma_{\text{ann}}$ to be one order of magnitude below the value given by eq.~\eqref{eq:sigma-area}, we get the conservative bound
\begin{equation}
  \Lambda \lesssim \SI{3}{TeV}.  
\end{equation}


\section{Conclusions}
\label{sec:conclusions}
\medskip

Skyrmions were originally introduced as topologically stable static field configurations in relativistic quantum field theory. Their purpose was to explain the existence of baryons in terms of topological solitons in an effective low-energy theory of mesons. In recent years, skyrmions of a different type have been experimentally observed in magnetic ordered materials. These magnetic skyrmions are described by a non-relativistic field theory on a discretised spin-lattice system. 
Importantly, these magnetic skyrmions are not protected by topology, they are separated only by finite energy barriers from the ground state.
While the experimentally observed existence of skyrmions in condensed matter systems has received a lot of attention, the theoretical investigation of skyrmions in particle physics, in particular the electroweak skyrmions in the presence of a dynamical Higgs field a  was lacking. 

\medskip

In this paper we showed that the interplay between a dynamical Higgs field and the electroweak gauge sector of the Standard Model leads to a non-trivial vacuum structure, that allows for the formation of electroweak skyrmions under rather generic circumstances. Like the skyrmions in condensed matter systems, the electroweak skyrmions are non-topological, they are not absolutely stable but have an exponentially long lifetime. We clarified the relation between the well-studied {\it electroweak sphalerons} that are saddle-points between vacua with different $n_\mathrm{CS}$ and the {\it skyrmionic sphalerons} that are saddle-points between skrymion field configurations with different $n_\mathrm{Sk} \neq 0$, including the trivial vacuum at $n_\mathrm{Sk} =0$. Electroweak skyrmions can unwind through highly-suppressed instanton processes and lead to striking signatures with $\Delta(B+L)=6\Delta n_\mathrm{CS}$. 

\medskip

We identified dimension-8 operators that stabilise the electroweak skyrmion as a spatially localised soliton field configuration with finite size. To assess how the energy of the skyrmion depends on the suppression scale of the effective operator $\Lambda$, i.e. the Skyrme term, we use a neural network to calculate the minimum of its energy functional. The mass of the electroweak skyrmion scales as $M_\mathrm{Sk} \simeq 0.35 \, {4 \pi v^3}/{\Lambda^2}$ and its radius as $R_\mathrm{Sk} \simeq 0.6\,{v}/{\Lambda^2}$.

\medskip

The dimension-8 Skryme term can be induced by a large class of UV models. We gave examples for minimal extensions of the Standard Model by spin-0, spin-1/2 and spin-1 particles which each individually, and as a subset of a more comprehensive extension of the Standard Model, would contribute to the emergence of a Skyrme term. The Skyrme term can also have a non-perturbative origin, as it had in the case of strong interactions or in the technicolour models.

\medskip

While the electroweak skyrmion production cross section is highly suppressed in collisions of nucleons or leptons, the LHC or future high-energy collider experiments provide a promising avenue to probe the Skyrme term in multi-Higgs-associated production processes. In turn, to be able to give experimentally measured final states an interpretation in terms of the presence of a Skryme term one needs to extend the SMEFT framework to operators of dimension-8 in global EFT analyses. Importantly,  electroweak skyrmions can be a viable dark matter candidate and, thus, provide a solution to the dark matter problem.

\medskip

For the electroweak skyrmion to be heavy and therefore less stable, the Skyrme term needs a small suppression scale. Here, we only considered weakly-coupled UV theories as the source of the Skryme term, for which a small $\Lambda$ is difficult to accommodate. However, the Skryme term could also be induced by a strongly-coupled dark sector, which would result naturally in a smaller suppression scale.

\medskip

Due to the profound implications the electroweak skyrmion can have on early Universe physics, dark matter and collider phenomenology, and in general on improving our understanding of the electroweak vacuum structure further experimental investigations seem not only warranted but required.


\bigskip

\section*{Acknowledgements}
We would like to thank Supratim Das Bakshi and Joydeep Chakrabortty for helpful discussions.


\appendix

\section{Skrymion energy calculation using a Neural Network}
\label{app:neural-nets}

To find the minimum energy for given values of the parameters $\kappa$, $\xi$ and $n$, we follow \cite{Piscopo:2019txs, Balaji:2020yrx} to model the set of functions $f_1$, $f_2$, $h$ and $b$ using a neural net with a single 30-unit layer. That is, we parametrize them as
\begin{equation}
  (f_1(r), f_2(r), b(r), h(r)) =
  \sum_{i=1}^{30} \left[
    \mathbf{b}^{(2)}_i
    + \frac{\mathbf{w}^{(2)}_{i}}{1 + \exp\left(-b^{(1)}_i - w^{(1)}_i r\right) }
  \right],
\end{equation}
where boldface is used to denote 4-component vectors of parameters. The net is trained using the Adam minimization algorithm, with loss function given by
\begin{equation}
  L[f_1, f_2, b, h] = E_{\text{nat}}[f_1, f_2, b, h]
  + \omega_{\text{BC}} \sum_k \text{BC}_k[f_1, f_2, b, h]^2
  + \omega_n \left(n_W[f_1, f_2, b, h] - n_W\right)^2.
\end{equation}
Here, $E_{\text{nat}}[f_1, f_2, b, h]$ is the energy of the configuration, as defined in eqs.~\eqref{eq:energy}, $\operatorname{BC}[f_1, f_2, b, h]$ is a tuple containing the differences between the values of the $f_1$, $f_2$, $b$ and $h$ functions at the boundaries and the values they are assigned by the boundary conditions, and $n[f_1, f_2, b, h]$ is the quantity defined in eq.~\eqref{eq:cs-number}. The integral for the energy is computed by averaging over 1000 equally distributed points from $r = 0$ to $r = 10$.  The weights $\omega_{\text{BC}}$ and $\omega_n$ need to be adjusted depending on the value of $\kappa$ and $\xi$. They should be such that minimizing $L$ amounts to minimizing the energy while satisfying the boundary and $n_W[f_1, f_2, b, h] = n_W$ conditions. This is achieved for $\omega_{\text{BC}} \simeq \omega_n \simeq 10^4$. Higher values ensure that the conditions are satisfied, but setting them as low as possible gives faster convergence of the training procedure.

The minimization algorithm is run until the relative improvement of the loss function is less than $10^{-5}$ over $1000$ epochs. It typically takes a few $\times 10^5$ epochs to reach this condition. To check the consistence of the results, we have repeated several instances of the same calculation for various values of the parameters obtaining discrepancies in the energy of the solutions that are less than $5\%$.

\bibliographystyle{JHEP}
\bibliography{references}

\end{document}